\documentclass[twocolumn]{article}  
\usepackage{graphicx}
\graphicspath{{./figures/}}
\usepackage{txfonts}
\usepackage{amssymb}
\usepackage[below]{placeins}
\usepackage[numbers]{natbib}
\usepackage{url}

\usepackage[a4paper,margin=15mm]{geometry}

\newcommand\inst[1]{$^{#1}$}
\renewcommand\and{ and }

\begin{document}

   \title{Spectroscopy of the stellar wind in the Cygnus X-1 system}

%   \subtitle{}

   \author{Ivica Mi\v{s}kovi\v{c}ov\'a\inst{1},
          Manfred Hanke\inst{1},\\
	  J\"orn Wilms\inst{1},
          Michael A. Nowak\inst{2},
          Katja Pottschmidt\inst{3,4},
          Norbert S. Schulz\inst{2}
	  }
          
   \date{\begin{minipage}{\textwidth}\normalsize
     \inst{1} Dr. Karl Remeis-Sternwarte, %Astronomisches Institut der
     Universit\"at Erlangen-N\"urnberg \& ECAP, Sternwartstr. 7, 96049
     Bamberg, Germany\\
 %     \and
     \inst{2} MIT Kavli Institute for Astrophysics and Space Research, NE80-6077, 77 Mass. Ave., Cambridge, MA 02139, USA\\
%     \and
     \inst{3} CRESST and NASA Goddard Space Flight Center,
     Astrophysics Science Division, Code 661, Greenbelt, MD 20771, USA\\
%     \and
     \inst{4} Center for Space-Science \& Technology, University of
     Maryland Baltimore County, 1000 Hilltop Circle, Baltimore, MD
     21250, USA\\~\\
 \centerline{\textbf{\large Abstract}}\\
   {The X-ray luminosity of black holes is produced through the accretion of
material from their companion stars. Depending on the mass of the donor
star, accretion of the material falling onto the black hole
through the inner Lagrange point of the system or accretion by the
strong stellar wind can occur.
   Cygnus X-1 is a high mass X-ray binary system, where the black hole is
powered by accretion of the stellar wind of its supergiant companion
star HDE226868. As the companion is close to filling its Roche lobe, the
wind is not symmetric, but strongly focused towards the black hole.
   Chandra-HETGS observations allow for an investigation of this focused
stellar wind, which is essential to understand the physics of the
accretion flow. We compare observations at the distinct orbital phases
of 0.0, 0.2, 0.5 and 0.75. These correspond to different lines of sights
towards the source, allowing us to probe the structure and the dynamics
of the wind.}
\end{minipage}
     }


%  \keywords{
%\begin{multicols}{1}
%\onecolumn[

   \maketitle
%

%  \abstract

%\end{multicols}

%\begin{multicols}{2}

\section{Introduction}

\subsection{Stellar winds of O stars}

Stellar winds of early (O or B) type stars are driven by the radiation pressure of copious absorption lines 
present in the ultraviolet part of the spectrum on material in the stellar atmosphere \citep{castor75}. 
Therefore the winds are very strong; common mass loss rates are $\sim$10$^{-6} M_{\odot}/\mathrm{year}$.
Since primaries of high-mass X-ray binaries are O or early B stars \citep{conti78}, which radiate in the UV, this radiation is strong 
enough to produce such a wind.
According to simulations of line-driven winds, perturbations are present and dense and cool 
inhomogeneities are created in the wind \citep{feldmeier97}. Larger density, velocity, and temperature variations compress the gas further, creating ``clumps''. Current knowledge about stellar winds assumes two disjunct components of O star winds: 
cool dense clumps and hot tenuous gas. Sako et al. \citep{sako02} showed that observed spectra of X-ray binaries can only be explained as originating from an environment, where the cool and dense clumps are embedded in the 
photoionized gas.

\subsection{Cygnus X-1}
Cygnus X-1 is a binary system where the X-ray source is a black hole \citep{bolton72, webster72}, and $\sim18\,M_\odot$ \citep{herrero95}, O 9.7 Iab type star HDE 226868 is its companion \citep{walborn73}. Stellar wind accretion plays a major role in the mass transfer process, because \mbox{Cyg X-1} belongs to the High-Mass X-ray Binaries (HMXB), which is in contrast to Low-mass X-ray Binaries (LMXB), where Roche lobe overflow is more important and accretion disk accretion occurs.
There are strong tidal interactions in the system. Moreover, the donor star fills $\sim$90\,\% of its Roche volume \citep{conti78,giesbolton86b}. Therefore the wind is not symmetric, but \emph{focused} towards the black hole \citep{friend82}, such that density and mass loss rate are higher along the binary axis. %(Fig. \ref{FriendCastor1982}).
The fact that such a high percentage of the Roche lobe is filled, however, means that we cannot exclude Roche lobe overflow taking place as well.

\subsection{Hard and soft state of Cygnus X-1}

\begin{figure}
\resizebox{\hsize}{!}{\includegraphics{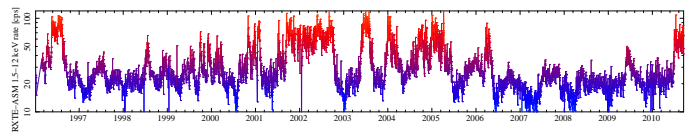}}
\caption{$1.5 - 12\, \mathrm{keV}$ light curve of Cyg X-1 obtained from 1996 to 2010 by the \textsl{RXTE}-ASM. Over time, transitions from hard to soft state occur. Compared to the time spent in the low-luminosity \emph{hard} state (blue, $\lesssim\:$40\,c/s for ASM countrate) Cyg X-1 spent less time in the high-luminosity \emph{soft} state (red, $\sim$100\,c/s for ASM countrate).
\label{RXTE-ASM_0}}
\end{figure}

Black hole binaries show two principal types of emission called the hard or soft state, which differ in the shape of the X-ray spectrum, the timing properties and the radio emission. Cyg X-1 spends most of the time in the hard state source with a hard, exponentially cut-off powerlaw spectrum, strong short term variability and steady radio emission \citep{pottschmidt03,wilms06}. However, transitions between hard and soft states are observed (Fig.~\ref{RXTE-ASM_0}).

How exactly the wind properties differ between states and what triggers state transitions are questions remaining to be answered. One possibility is that states correspond to different configurations of the accretion flow \citep{smith02}, which cause differences in energy dissipation. Another possibility is that changes of the wind properties themselves trigger state transitions \citep{gies03}.
The mass transfer process in either case (HMXB or LMXB) provides extremely efficient energy release and produces luminosities of $\sim$ $10^{37}$ $\mathrm{erg/s}$ in general. For such a luminosity, which is typical for Cyg X-1, the X-ray source produces a considerable feedback on the wind by photoionization of its nearby environment \citep{blondin94}, which contributes to the complex wind structure.

\section{Observations and data analysis}

\subsection{Observations and orbital coverage}

The spectra used for our analysis were obtained by the High Energy Transmission Grating Spectrometer (HETGS -- HETG in combination with ACIS, Advanced CCD Imaging Spectrometer) on board the \textsl{Chandra} observatory. \textsl{Chandra} ACIS observations are performed in two different modes: timed exposure (TE) mode and continuous clocking (CC) mode. In TE mode, the CCD is exposed for some time and then its data are transfered to the frame store, which is read out during the next exposure. The readout time required for the full frame store is 3.2s. In CC mode, the columns are read out continuously, which reduces the readout time to 3ms \citep{garmire03}\footnote{Chandra X-ray Center, The Chandra Proposers' Observatory Guide, 2009, \url{http://cxc.harvard.edu/proposer/POG/}}. When the source is very bright, more than one photon may reach the same pixel in one frame time (pile-up). These photons are misinterpreted as one single event with higher energy. CC mode is usually used to avoid pile-up. 

High-resolution spectra of persistently bright sources like Cyg X-1 provide the unique possibility of probing the structure of the wind \emph{directly}.
However, this structure and therefore also the properties of the wind (density, velocity, ionization state) change with different lines of sight, which correspond to the different orbital phases. Thus, a good coverage of the binary orbit is desirable.

\begin{figure}% if figure*:in two column paper, both of the columns are used for figures
\begin{minipage}[b]{.4\columnwidth}
a)
\includegraphics[width=\columnwidth,angle=0.]{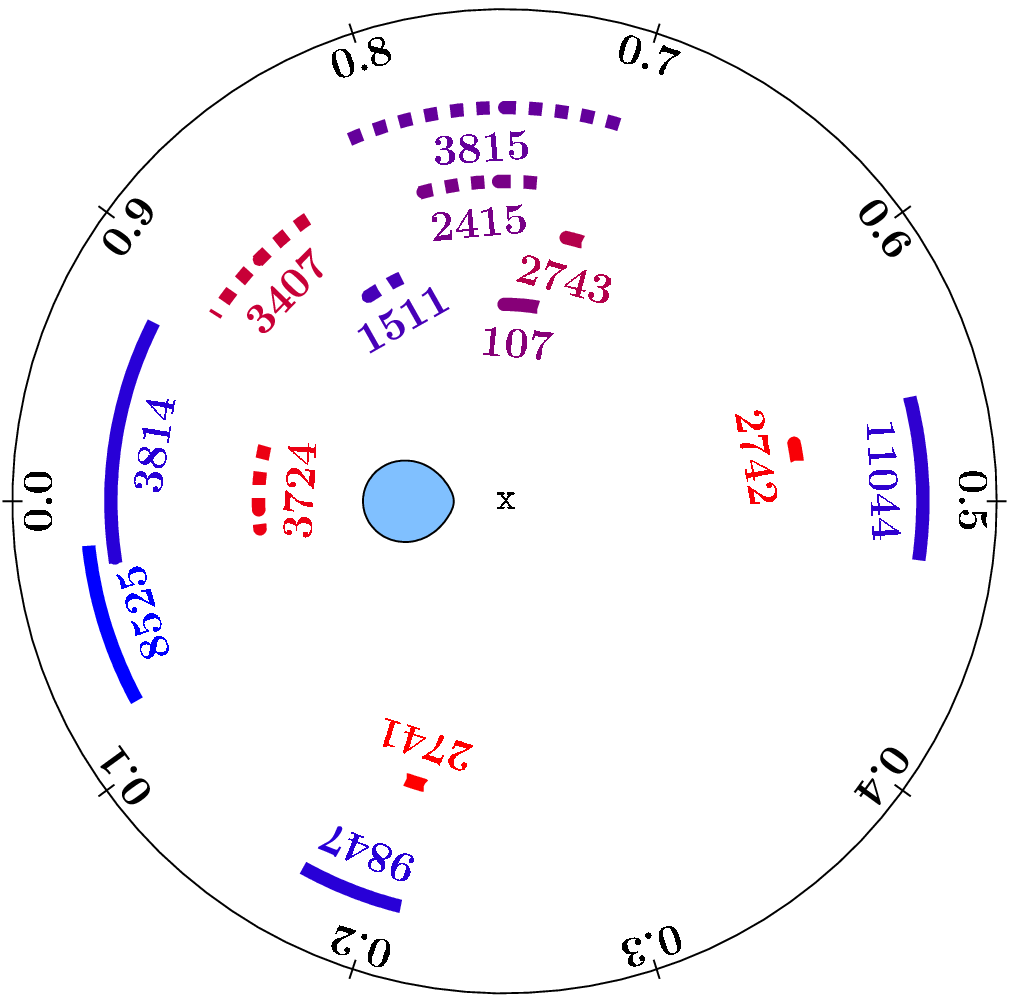}
\end{minipage}
\begin{minipage}[b]{.15\columnwidth}
b)
\includegraphics[width=\columnwidth,angle=0.]{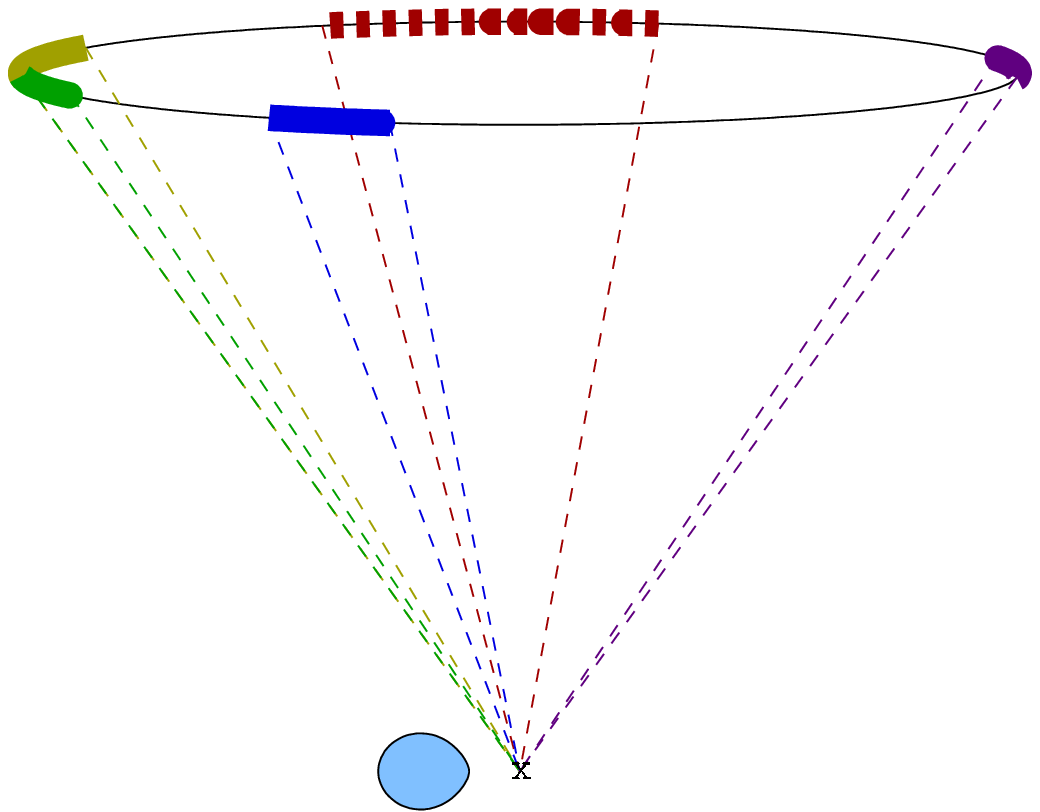}
\end{minipage}
\begin{minipage}[b]{.4\columnwidth}
c)
\includegraphics[width=\columnwidth,angle=0.]{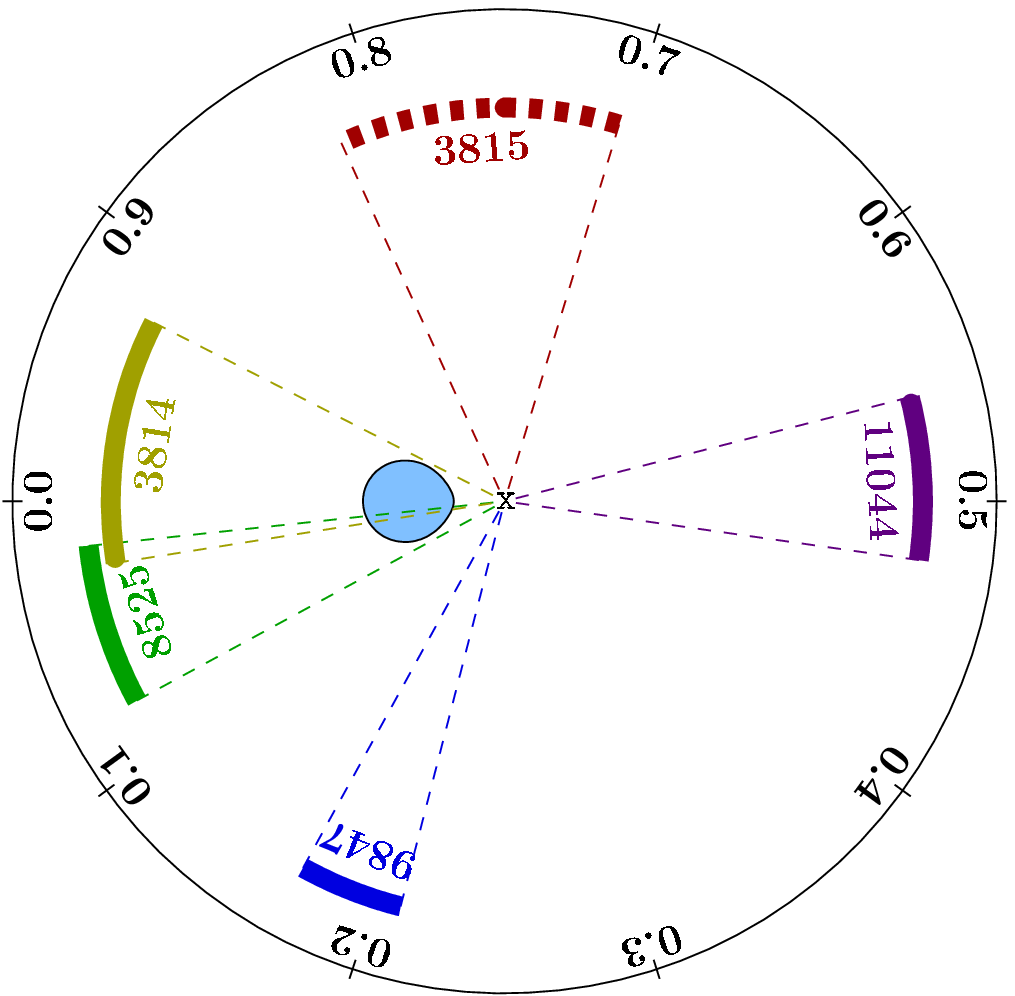}
\end{minipage}
\caption{a) Polar view of the Cyg X-1 orbit and illustration of observation coverage with \textsl{Chandra}. Before 2010, January, there were 13 observations available, mostly covering the part of the orbit around phase $\phi{\approx}$0. Another observation (ObsID 11044) was obtained at $\phi{\approx}$0.5. A short observation (ObsID 2742) was the only one at this phase before, but since it was obtained in TE mode during the soft state of the source, it strongly suffered from pile-up.
Full lines (dashed lines) display TE mode (CC mode) observations. Changes from blue to red colour correspond to changes from hard to soft state.
b)~Colorcoded orbital phases corresponding to lines of sight towards Cyg X-1 taking into account an inclination of $\sim$ 35$^{\circ}$.
c)~Highlighted observations at phases of $\phi{\approx}$0.0 (ObsID 3814 and ObsID 8525), 0.2 (ObsID 9847), 0.5 (ObsID 11044) and 0.75 (ObsID 3815), which are analyzed and compared in this work.      
\label{Chandra_coverage}}
\end{figure}

Observations (whether in the hard or soft state) which are currently available cover part of the orbit around phase $0.0$, between phases $0.7$ and $0.2$, and around phase $0.5$, with the latter only obtained in January, 2010 (Fig. \ref{Chandra_coverage}a). We focus here on the comparison of observations obtained at four distinct phases of $\phi=0.0$, which is defined at the time of superior conjunction of the black hole (ObsID 3814 and ObsID 8525), $0.2$ (ObsID 9847), $ 0.5$ (ObsID 11044) and $0.75$ (ObsID 3815), see Fig. \ref{Chandra_coverage}c.
The latter was obtained in CC mode, while all others were obtained in TE mode. This difference has no influence on our comparison. While the calibration of CC mode does not allow for an adequate modelling of the whole continuum shape, local absorption lines, which are our primary interest, are not affected.

The phase $0.0$ coverage is extremely important, since due to the inclination of $\sim$ 35$^{\circ}$ of the Cyg X-1 orbital plane, it corresponds to looking through the densest part of the wind close to the stellar surface (Fig. \ref{Chandra_coverage}b). The distribution of X-ray dips with orbital phase peaks around phase $0.0$ \citep{balucinska00}. The observation around phase $0.5$ provides a great opportunity to close a gap in defining the general picture of the wind structure.
%Light curves obtained for all selected observations are shown in Fig. \ref{Chandra_lightcurves}.

While all recent \textsl{Chandra} observations caught Cyg\,X-1 in the hard state at $\lesssim\:$100\,c/s (of \textsl{Chandra} countrate), comparable to the observation at $\phi{\approx}$0 \citep{hanke09}, the spectrum was softer and the flux was more than twice as high during the observation at $\phi{\approx}$0.7. The light curves at $\phi{\approx}$0 are modulated by strong and complex \emph{absorption\, dips}, but dipping occurs already at $\phi{\approx}$0.7 and has not ceased at $\phi{\approx}$0.2, though the dip events seem to become shorter with distance from $\phi{=}$0. The light curve at $\phi{\approx}$0.5 is totally \emph{free of dips}, yielding 30\,ks of remarkably constant flux.

\subsection{Absorption dips}

\begin{figure}
\resizebox{\hsize}{!}{\includegraphics{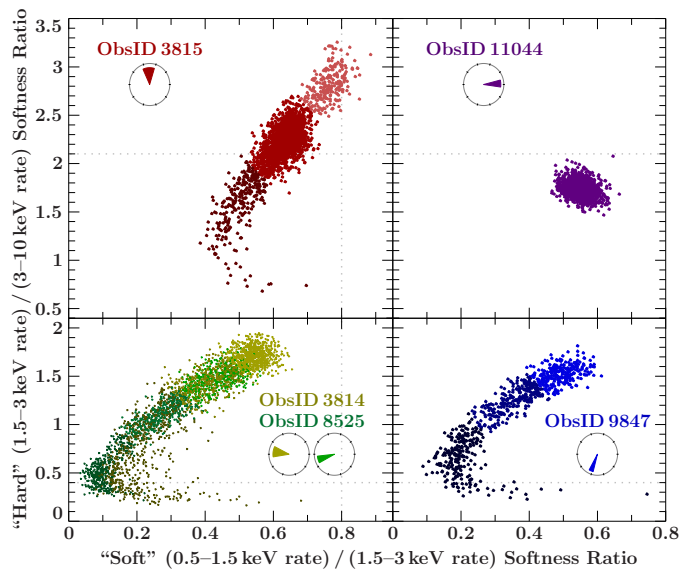}}
\caption{ All of these diagrams show a ``soft softness ratio" on the $x$- and a ``hard softness ratio" on the $y$-axis. Dipping produces a clear track in the color-color diagrams:
Both colors harden towards the lower left corner, due to increased absorption. However, during extreme dips, the soft color becomes softer again, which is likely due to partial coverage.
\label{color-color}}
\end{figure}

According to general assumption, absorption dips, during which the soft X-ray flux decreases sharply, originate from inhomogeneities --``clumps''-- present in the wind, where the material is of higher density and lower temperature \citep{castor75,sako02}. According to the softness ratios in the color-color diagram (Fig. \ref{color-color}), different stages of dipping can be classified.

\begin{figure}
\resizebox{\hsize}{!}{\includegraphics{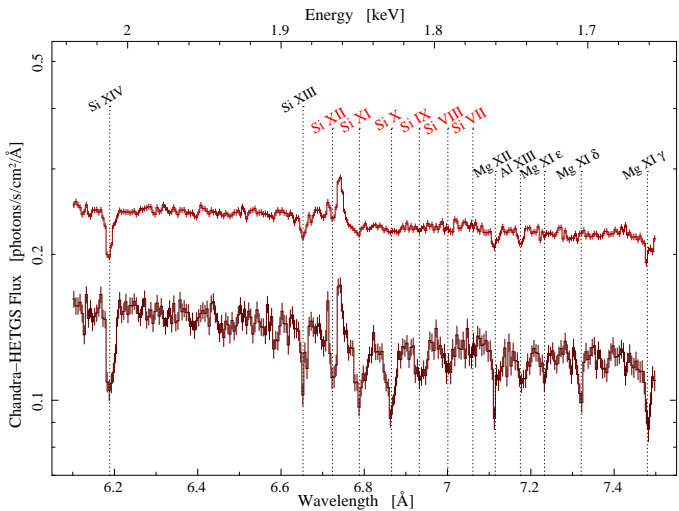}}
\caption{ ``Dip'' and ``non-dip'' spectra from the observation at phase $\phi{\approx}$0.7 are shown for comparison of absorption lines present in different stages of dipping. The \emph{reduction in flux} in the spectra is real and due to dips.
In the "non-dip" spectrum, only absorption lines of Si\,\textsc{xiv} and Si\,\textsc{xiii} are present, while in the "dip" spectrum the whole series of Si\,\textsc{xii}--\textsc{vii} appears.
\label{flux_Si-region}}
\end{figure}

Figure~\ref{flux_Si-region} shows the spectrum from the observation at phase $\phi{\approx}$0.7 (ObsID\,3815) split into "dip" and "non-dip" stages in the wavelength interval of the Si-region between 6\,\AA\, and 7.5\,\AA. 
While absorption lines of Si\,\textsc{xiv} and Si\,\textsc{xiii} are already present in the non-dip spectrum, the dip spectra contain \emph{additional strong absorption lines} that can be identified with K$\alpha$ transitions of lower ionized Si\,\textsc{xii}--\textsc{vii}.
The strength of the low-ionization lines increases with the degree of dipping, indicating that the latter is related to clumps of lower temperature. Moreover, the clumps are of higher density than their surroundings. 
In 2008 our group organized a multi-satellite observational campaign with \textsl{XMM-Newton},\textsl{Chandra} (ObsID 8525 and ObsID 9847), \textsl{Suzaku}, \textsl{RXTE}, \textsl{INTEGRAL} and \textsl{Swift} observing Cyg X-1 simultaneously. Dips shortly after phase $\phi{\approx}$0 were so strong, that they were seen by all the instruments involved in the campaign, even \textsl{RXTE}--PCA or \textsl{INTEGRAL}--ISGRI \citep{hanke08}. 

\begin{figure}[!h]
\resizebox{\hsize}{!}{\includegraphics{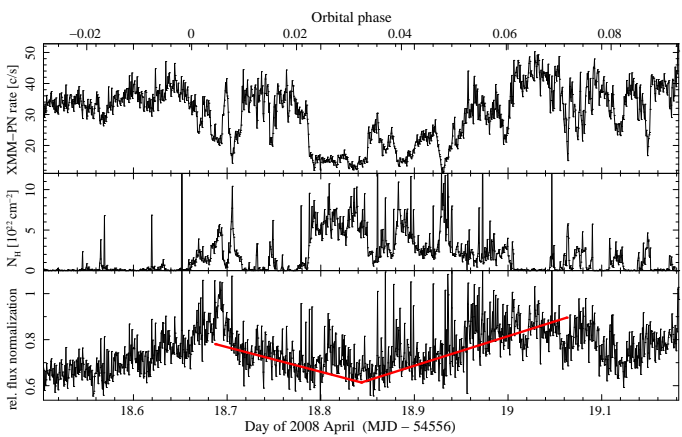}}
\caption{Absorption dips and scattering of X-rays as seen with \textsl{XMM-Newton}, EPIC-pn.
a) The light curve in the energy band $0.3 - 10\, \mathrm{keV}$ shows absorption dips, which are identical to dips observed by \textsl{Chandra} shortly after $\phi{\approx}0$.
b) Hydrogen column density of the neutral absorption model increases strongly when dips occur.
c) Relative flux normalization constant, which is consistent with the scattering trough seen in hard X-rays \citep{hanke10}.
\label{XMM-EPIC-pn}}
\end{figure}

The light curve from XMM-Newton is shown in Fig.~\ref{XMM-EPIC-pn} \citep{hanke10}. Where the dips occur in the light curve, the hydrogen column density, $N_{\mathrm{H}}$, increases strongly.

As shown in the third panel, however it is not only the pure absorption which causes the dips. Thomson scattering contributes during these times, causing longer time scale variations, also at hard X-rays. A possible explanation is in the existence of dense and (nearly) neutral clumps (causing the sharp dips) embedded in ionized halos (causing the scattering) \citep{hanke10}.

\subsection{Spectroscopy}

\begin{figure}
\resizebox{\hsize}{!}{\includegraphics{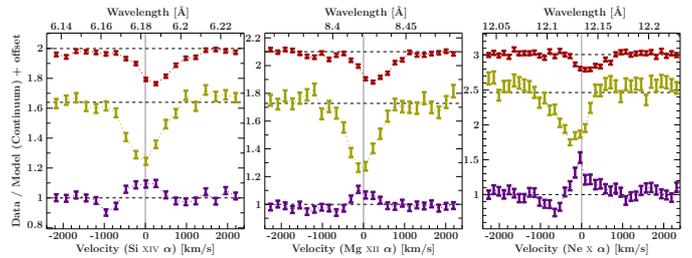}}
\caption{ Absorption and P Cygni profiles of Si\,\textsc{xiv}, Mg\,\textsc{xii} and Ne\,\textsc{x}.
While the observations at $\phi{\approx}0$ and $\phi{\approx}0.75$ show clear absorption profiles, although redshifted for $\phi{\approx}0.75$, the observation at $\phi{\approx}0.5$ shows emission at $v_{\rm rad}{\approx}0$ and blueshifted absorption.
\label{H-lineProfiles}}
\end{figure}

We separate the ``non-dip" and the ``dip" parts of the observations. The ``non-dip" spectrum is extracted from the least absorbed phases at the upper right corner of the color-color diagram (except for ObsID\,3815) and the spectroscopic results here refer to these ``non-dip" phases.
The highly photoionized wind is detected at $\phi{\approx}0$ via numerous strong absorption lines at $v_{\rm rad}{\approx}0$ (Fig. \ref{H-lineProfiles}). 

The lack of appreciable Doppler shifts can be explained by the wind flow being orthogonal to the line of sight. %at the location where the absorbing ionization stages dominate in the Str\"omgren zone where the absorbing ions dominate the ionization balance\citep{hanke08},\citep{hanke09}.
In contrast, the recent observation (ObsID 11044) at $\phi{\approx}0.5$ reveals \emph{for the first time for Cyg X-1} clear \emph{P\,Cygni profiles} with a strong emission component at a projected velocity $v_{\rm rad}{\approx}0$, while the weak absorption components occur at a blueshift of $\approx$500--1\,000\,km/s.
%If, despite the rotated line of sight by $2i{\approx}70^\circ$, the emission at $\phi{=}0.5$ is caused by the same plasma as the absorption at $\phi{=}0$,
If we observe the same plasma in both cases, this indicates that the real velocity must be small, i.e., we are probing a dense, low-velocity wind close to the stellar surface.
The fact that the absorption line profiles measured at $\phi{\approx}0.75$ are redshifted by $\approx$200--300\,km/s indicates that the wind flow is not radial from the star, as a radial wind (i.e., directing away from the star) would always give a blueshifted velocity when projected onto the line of sight at phases $\phi=0.25 - 0.75$.

\section{Summary}

The new Chandra observation of Cygnus X-1 at orbital phase 0.5 obtained in January 2010 allows us to compare observations at the four distinct orbital phases 0.0, 0.2, 0.5 and 0.75. With such a coverage, the full structure of the wind starts to reveal itself. At phase 0.0 we look through the densest part of the wind, as it is focused towards the black hole. The light curve is modulated by strong absorption dips. The flux decreases strongly during such dips, consistent with being caused by dense and cool clumps of material embedded in the more ionized wind. 
While absorption lines of Si\,\textsc{xiv} and Si\,\textsc{xiii} are already present in the non-dip spectrum, in the dip spectra also K$\alpha$ transitions of lower ionized Si appear, whereas the strength of these lines increases with the degree of dipping.
An especially interesting result is the totally flat light curve around phase 0.5. While dipping has started around phase 0.7, is the strongest around 0.0 and still present at 0.2, it has vanished at 0.5. We therefore proposed for the next observation between phases 0.25 and 0.4 to investigate the transition between dipping and non-dipping phases. 
Spectroscopic analysis showed another interesting result. In the spectrum at phase 0.5, clear P-Cygni profiles of Lyman $\alpha$ transitions were observed for the first time for Cyg X-1. We observe here strong emission components at a projected velocity $v_{rad}{\approx}0$ in contrast to pure absorption observed at phase 0.0.
Detailed modeling of photoionization and wind structure is in progress.

%\begin{acknowledgements}
\paragraph{Acknowledgements}
 The research leading to these results was funded by the European Community's Seventh Framework Programme (FP7/2007-2013) under grant agreement number ITN 215212 "Black Hole Universe" and by the Bundesministerium f\"ur Wirtschaft und Technologie under grant number DLR 50 OR 0701.
%\end{acknowledgements}

%\bibliographystyle{aa}
%\bibliographystyle{plain}
%\bibliographystyle{PoS}
%\bibliography{mnemonic,cygnusx1_noJournalBackslash}

%\end{multicols}

\end{document}